\documentclass[12pt]{article}
\usepackage{sw20lart}

\input tcilatex
\QQQ{Language}{
American English
}

\begin{document}

\title{Non-Static Spherically Symmetric Solution of Einstein Vacuum Field Equations
With $\Lambda $}
\author{Amir H. Abbassi \\
Department of Physics, School of Sciences, \\
Tarbiat Modarres University, P.O.Box14155-4838,\\
Tehran,Iran.\\
E-mail: ahabbasi@net1cs.modares.ac.ir}
\maketitle

\begin{abstract}
The Schwarschild-de Sitter metric is the known solution of Einstein field
equations with cosmological constant term for vacuum spherically symmetric
space around a point mass $M$. Recently it has been reported that in a $%
\Lambda $-dominant world the Schwarzschild type coordinate systems are
disqualified by redshift-magnitude test as a proper frame of
reference(gr-qc/9812092). We derive the solution in a FRW type coordinate
system which is qualified according to the mentioned test. Asymptotically it
approachs to the non-static form of deSitter metric. The obtained metric is
transformable to Schwarzschild-deSitter metric. It is an analytic function
of $r$ for all values except $r=0$ which is singular. This is carried out
with no making use of Eddington-Finkelstein coordinates and without entering
any cross term in the metric.

\bigskip\ 

\end{abstract}

\newpage\ \ Today the best standard candles known in cosmology probably are
type Ia supernovae. They have became the principle indicator for the
determination of distance in the Hubble flow and consequently measuring more
accurately the Hubble constant and the cosmological deceleration. The
startling result of the recent high-z supernovae observation is that the
deceleration parameter comes out negative, $q_0\approx -\frac 12.$ This
implies an acceleration expansion rate. The following relation between $q_0$%
, $\Omega _M$ and $\Omega _\Lambda $ holds

\begin{equation}
q_0=\frac{\Omega _M}2-\Omega _\Lambda  \label{1}
\end{equation}
which the values of $\Omega _M\approx 0.3$ and $\Omega _\Lambda \approx 0.7$
are favored by new achievements. A positive cosmological constant is
inferred from recent measurements and indicate that cosmological constant is
the dominant term in the Friedmann equation at our epoch[1,2,3,4,5]. So far
in standard models of gravitating systems the existence of cosmological
constant have been simply ignored,but after these observations the
theoretical models should be modified to include this term as a part of
reality.

A theoretical issue as has been well explained in standard reviews of
cosmological constant is that the theoretical expectation for $\Lambda $
exceeds its observational value by 120 orders of magnitude[6,7]. Meanwhile
it worth to mention that there is some way to resolve this deficiency. It is
proved that there exists a different commutation relation between space
coordinate and its momentum for massless particles[8]. The field theoretical
counterpart of this new commutation relation leads to zero vacuum energy[9].

The familiar solution of the Einstein field equations with cosmological
constant for a vacuum space around a center of spherical symmetry(SS) is
given by the Schwarzschild-deSitter metric[10]. It has the following form ($%
G=c=1$):

\begin{equation}
ds^2=(1-\frac{2M}r-\frac \Lambda 3r^2)dt^2-(1-\frac{2M}r-\frac \Lambda
3r^2)^{-1}dr^2-r^2(d\theta ^2+\sin ^2\theta \;d\varphi ^2)  \label{2}
\end{equation}
where $\Lambda $ stands for cosmological constant and $M$ is the mass of an
individual point mass which is the source of SS. This metric has two
coordinate type singularity at distance very close to $2M-\frac{M^3\Lambda }%
2 $ and $\sqrt{\frac 3\Lambda }-M$ and an intrinsic singularity at $r=0$. On
the geodesic with $d\theta =d\varphi =0\;,\;ds^2$ does change sign in
passing from $2M-\frac{M^3\Lambda }3<r<\sqrt{\frac 3\Lambda }-M$ interval to
the neighboring intervals $r<2M-\frac{M^3\Lambda }3$ and $r>\sqrt{\frac
3\Lambda }-M.$ This enforces us to give up our common sense and accept the
ad hoc character exchange of $r$ and $t$ as space and time coordinate. On
the other hand according to the principle of causality we expect when $M$
the source of SS tends to zero the symmetry of the metric which is the
effect of the existence of $M$ disappear, but in Schwarzschild-deSitter
metric the contrast of this happens and SS remains. The crucial deficiency
of the Schwarzschild-deSitter metric is that in a $\Lambda -$dominated
universe the Schwarzschild type comoving coordinate system gives a
redshift-magnitude relation which does not confirm with observational
data[11]. This discards the Schwarzschild type coordinate as a proper frame
of reference in the presence of $\Lambda .$ A FRW type comoving coordinate
system gives a correct prediction for redshift-magnitude relation which
quite well agrees with observation.

We are going to find the solution of Einstein field equations with
cosmological constant for a vacuum spherically symmetric space in a proper
FRW type coordinate system. Although a system consisting a point mass $M$
possesses SS, it is plausible to expect that at very large scales compared
to $M$ the space looks homogenous and isotropic within good approximaton and
its related metric to be deSitter metric in its FRW form. Thus it will not
restrict the generality of the problem if take the following form for $ds^2,$

\begin{equation}
ds^2=B(r,t)dt^2-R^2(t)\;[A(r,t)dr^2+r^2(d\theta ^2+\sin ^2\theta \;d\varphi
^2)]  \label{3}
\end{equation}
where $A$ and $B$ tends to one in the large distance limit and $R(t)$ the
scale factor satisfies;

\begin{equation}
\left( \frac{\dot{R}}R\right) ^2=\frac{\ddot{R}}R=\frac \Lambda 3  \label{4}
\end{equation}
Taking $ds^2=-g_{\mu \nu }\;dx^\mu dx^\nu $, the nonvanishing components of $%
\;g_{\mu \nu }$ for metric (3) are:

\begin{equation}
g_{rr}=R^2A\;\;,\;\;g_{\theta \theta }=R^2r^2\;\;,\;\;g_{\varphi \varphi
}=R^2r^2\sin ^2\theta \;\;,\;\;g_{tt}=-B  \label{5}
\end{equation}
Since the off diagonal components of the metric are zero, its contravariant
components are simply the inverse of the related covariant one.

\newpage\ Then the nonvanishing components of affine connection are:

\begin{eqnarray}
\Gamma _{rr}^r &=&\frac{\acute{A}}{2A}\;\;\;,\;\;\;\Gamma _{rt}^r=\frac{\dot{%
R}}R+\frac{\dot{A}}{2A}\;,\;\Gamma _{\theta \theta }^r=-\frac rA\;,\;\Gamma
_{\varphi \varphi }^r=-\frac{r\sin ^2\theta }A\;,\;\Gamma _{tt}^r=\frac{%
\acute{B}}{2R^2A}  \nonumber  \label{6} \\[0.01in]
\Gamma _{r\theta }^\theta &=&\frac 1r\;\;\;\;\;,\;\;\;\Gamma _{t\theta
}^\theta =\frac{\dot{R}}R\;\;\;\;\;\;\;\;,\;\Gamma _{\varphi \varphi
}^\theta =-\frac{\sin 2\theta }2  \nonumber \\
\Gamma _{r\varphi }^\varphi &=&\frac 1r\;\;\;\;\;,\;\;\;\Gamma _{\theta
\varphi }^\varphi =\cot \theta \;\;\;\;\;\;,\;\Gamma _{t\varphi }^\varphi =%
\frac{\dot{R}}R  \label{6} \\
\Gamma _{rr}^t &=&\frac{2\dot{R}RA+R^2\dot{A}}{2B}\;,\;\Gamma _{tr}^t=\frac{%
\acute{B}}{2B}\;,\;\Gamma _{\theta \theta }^t=\frac{r^2\dot{R}R}B\;,\;\Gamma
_{\varphi \varphi }^t=\frac{r^2\sin ^2\theta \;\dot{R}R}B\;,\;\Gamma _{tt}^t=%
\frac{\dot{B}}{2B}\;\;  \nonumber  \label{6}
\end{eqnarray}
where($^{\prime }$) and ($\cdot $) stand for the derivative with respect to $%
r$ and $t$ respectively. The nonvanishing components of Ricci tensor in
terms of the above affine connections are:

\begin{eqnarray}
R_{rr} &=&\frac{B^{^{\prime \prime }}}{2B}-\frac{\acute{A}}{rA}-\frac{\acute{%
B}}{4B}(\frac{\acute{B}}B+\frac{\acute{A}}A)-\frac{R^2A}B[\frac{\ddot{R}}R+%
\frac{2\dot{R}^2}{R^2}+\frac{2\dot{R}\dot{A}}{RA}+\frac{\ddot{A}}{2A} 
\nonumber \\
&&-\frac{\dot{R}\dot{B}}{2RB}-\frac{\dot{A}\dot{B}}{4AB}-(\frac{\dot{A}}{2A}%
)^2]  \label{7}
\end{eqnarray}

\ 

\begin{eqnarray}
R_{tt} &=&\frac B{R^2A}[-\frac{B^{^{\prime \prime }}}{2B}+\frac{\acute{B}}{4B%
}(\frac{\acute{B}}B+\frac{\acute{A}}A)-\frac{\acute{B}}{rB}]+\frac{3\ddot{R}}%
R+\frac{\ddot{A}}{2A}-\frac{\dot{A}^2}{4A^2}+\frac{\dot{R}\dot{A}}{RA} 
\nonumber  \label{8} \\
&&-\frac{3\dot{R}\dot{B}}{2RB}-\frac{\dot{A}\dot{B}}{4AB}  \label{8}
\end{eqnarray}

\begin{equation}
R_{\theta \theta }=-1+\frac 1A-\frac{r\acute{A}}{2A^2}+\frac{r\acute{B}}{2AB}%
+\frac{R^2r^2}B[-\frac{2\dot{R}^2}{R^2}-\frac{\ddot{R}}R+\frac{\dot{R}\dot{B}%
}{2RB}-\frac{\dot{R}\dot{A}}{2RA}]  \label{9}
\end{equation}

\[
R_{\varphi \varphi }=\sin ^2\theta \;R_{\theta \theta
}\;\;\;\;\;\;\;\;\;\;\;\;\;\;\;\;\;\;\;\;\;\;\;\;\;\;\;\;\;\;\;\;\;\;\;\;\;%
\;\;\;\;\;\;\;\;\;\;\;\;\;\;\;\;\;\;\;\; 
\]

\begin{equation}
R_{rt}=-\frac{\dot{A}}{Ar}-\frac{\acute{B}\dot{R}}{BR}\;\;\;\;\;\;\;\;\;\;\;%
\;\;\;\;\;\;\;\;\;\;\;\;\;\;\;\;\;\;\;\;\;\;\;\;\;\;\;\;\;\;\;\;\;\;\;\;\;\;%
\;\;\;\;\;  \label{10}
\end{equation}

The Einstein field equations with $\Lambda $ for empty space are reduced to

\begin{equation}
R_{\mu \nu }=-\Lambda \;g_{\mu \nu }  \label{11}
\end{equation}
With the use of (3) and (11) we may write

\begin{equation}
\frac{R_{rr}}{R^2A}+\frac{R_{tt}}B=0  \label{12}
\end{equation}
Now inserting (7) and (8) in (12) results

\begin{equation}
-\frac 1{R^2rA}(\frac{\acute{A}}A+\frac{\acute{B}}B)-\frac 1B(-\frac{2\ddot{R%
}}R+\frac{2\dot{R}^2}{R^2}+\frac{\dot{R}\dot{A}}{RA}+\frac{\dot{R}\dot{B}}{RB%
})=0  \label{13}
\end{equation}
Using (4) , (13) leads to

\begin{equation}
\frac 1{R^2rA}(\frac{\acute{A}}A+\frac{\acute{B}}B)+\frac{\dot{R}}{RB}(\frac{%
\dot{A}}A+\frac{\dot{B}}B)=0  \label{14}
\end{equation}
Let us define $\rho $ to be $\rho =R(t)\;r$ and assume

\begin{equation}
A(r,t)=A(\rho )\;\;\;\;\&\;\;\;\;B(r,t)=B(\rho )  \label{15}
\end{equation}
If (*) stands for the derivative with respect to $\rho $ then the partial
derivative of A and B with respect to $r$ and $t$ are

\[
\acute{A}=R(t)\;A^{*}\;\;\;,\;\;\;\dot{A}=\dot{R}(t)rA^{*} 
\]

\begin{equation}
\acute{B}=R(t)\;B^{*}\;\;\;,\;\;\;\dot{B}=\dot{R}(t)rB^{*}  \label{16}
\end{equation}
We may apply (16) to simplify (14) so it becomes

\begin{equation}
(\frac 1{rAR}+\frac{r\dot{R}^2}{RB})(\frac{A^{*}}A+\frac{B^{*}}B)=0
\label{17}
\end{equation}
Equation (17) holds if we have

\begin{equation}
\frac{A^{*}}A+\frac{B^{*}}B=0  \label{18}
\end{equation}
Integration of (18) with respect to $\rho $ gives $AB=$constant. since $A$
and $B$ need to become one at large distances, thus we have

\begin{equation}
A=B^{-1}  \label{19}
\end{equation}
This selection automatically satisfies the $rt$-component of the field
equation and may be considered as a justification for the assumption in
(15). The $\theta \theta $-component of the field equation together with (4)
and (19) yield

\begin{equation}
(-1+\frac 1A)-\frac{r\acute{A}}{A^2}-\frac{\Lambda R^2r^2}B-\frac{R\dot{R}r^2%
\dot{A}}{BA}=-\Lambda R^2r^2  \label{20}
\end{equation}
and inserting (16) and (19) in (20) gives

\[
(-1+\frac 1A)-\frac{\rho A^{*}}{A^2}-\Lambda \rho ^2(A-1)-\frac \Lambda
3\rho ^3A^{*}=0 
\]
or 
\begin{equation}
-1+\frac d{d\rho }(\frac \rho A)-\frac \Lambda 3\frac d{d\rho }[\rho
^3(\Lambda -1)]=0  \label{21}
\end{equation}
Integration of (21) with respect to $\rho $ results

\begin{equation}
\frac \rho A-\frac \Lambda 3\rho ^3(\Lambda -1)-\rho =\text{constant}\equiv C
\label{22}
\end{equation}
If we define $D\equiv \rho A$, then (22) in terms of $D$ becomes a quadratic
equation

\begin{equation}
-\frac \Lambda 3\rho D^2-(1-\frac \Lambda 3\rho ^2+\frac C\rho )D+\rho =0
\label{23}
\end{equation}
This quadratic equation has two real solutions:

\begin{equation}
D_{\pm }=\frac{(1-\frac \Lambda 3\rho ^2+\frac C\rho )\pm \sqrt{(1-\frac
\Lambda 3\rho ^2+\frac C\rho )^2+\frac{4\Lambda \rho ^2}3}}{-2\frac \Lambda
3\rho }  \label{24}
\end{equation}
By evaluating the post Newtonian limits we observe that merely $D_{-}$ is
physically acceptable and the actual value of $C$ is $-2M.$ Thus we have

\begin{equation}
B=A^{-1}=\frac{2\Lambda \rho ^2}3[\sqrt{(1-\frac{2M}\rho -\frac \Lambda
3\rho ^2)^2+\frac{4\Lambda }3\rho ^2}-(1-\frac{2M}\rho -\frac \Lambda 3\rho
^2)]^{-1}  \label{25}
\end{equation}
Eq.(25) may be written in a more elegant form

\begin{equation}
B=A^{-1}=\frac 12[\sqrt{(1-\frac{2M}\rho -\frac \Lambda 3\rho ^2)^2+\frac{%
4\Lambda }3\rho ^2}+(1-\frac{2M}\rho -\frac \Lambda 3\rho ^2)]  \label{26}
\end{equation}
Thus the final non-static form of $ds^2$ in a FRW reference frame is

\begin{eqnarray}
ds^2 &=&\frac 12[\sqrt{(1-\frac{2M}\rho -\frac \Lambda 3\rho ^2)^2+\frac{%
4\Lambda }3\rho ^2}+(1-\frac{2M}\rho -\frac \Lambda 3\rho ^2)]dt^2  \nonumber
\label{27} \\
&&-2e^{2\sqrt{\frac \Lambda 3}t}[\sqrt{(1-\frac{2M}\rho -\frac \Lambda 3\rho
^2)^2+\frac{4\Lambda }3\rho ^2}+(1-\frac{2M}\rho -\frac \Lambda 3\rho
^2)]^{-1}dr^2  \nonumber \\
&&-\rho ^2(d\theta ^2+\sin ^2\theta \;d\varphi ^2)  \nonumber \\
\rho &\equiv &e^{(\sqrt{\frac \Lambda 3}t)}r  \label{27}
\end{eqnarray}

We may summarize the functional behavior of the obtained metric as follows.
For the interval $\rho \gg 2M$ we may neglect the $\frac{2M}\rho $ term
comparing to others and consequently come to $B=A^{-1}=1.$ This means that
at these very large scales the essential symmetry of space restores. This
property also guarantees that when $M$ tends to zero the homogeneity and
isotropy of space restore. For scales not too large but comparable to $2M$ ,
(26) reduces to $B=A^{-1}=(1-\frac{2M}\rho -\frac \Lambda 3\rho ^2)$ which
exactly coincide with Schwarzschild-deSitter metric. For the interval $\rho $
quite smaller than $2M$ we have

\[
B=A^{-1}\approx \mid 1-\frac{2M}\rho -\frac \Lambda 3\rho ^2\mid [1+\frac{%
\frac{2\Lambda }3\rho ^2}{(1-\frac{2M}\rho -\frac \Lambda 3\rho ^2)^2}+(1-%
\frac{2M}\rho -\frac \Lambda 3\rho ^2)]/2 
\]

\begin{equation}
\approx \frac{\Lambda \rho ^3}{6M}\;\;\;\;\;\;\;\;\;\;  \label{28}
\end{equation}

The obtained metric does not change sign and has quite different behavior in
the intervals which Schwarzschild-deSitter metric faces with problems. In
contrast to Schwarzschild-deSitter metric that at origin gives $%
g_{tt}=+\infty $ and $g_{rr}=0$, this new metric gives $g_{tt}=0$ and $%
g_{rr}=+\infty $. thus the non-static metric is analytic everywhere except
at $\rho =0$. At $\;\rho =2M-\frac{M^3\Lambda }3$ that is the familiar
horizon of black holes the metric gives

\begin{equation}
A(2M-\frac{M^3\Lambda }3)\approx \sqrt{\frac \Lambda 3}\times 2M  \label{29}
\end{equation}
On geodesic with $d\theta =d\varphi =0$, the relation between $\;dt$ and $dr$
at this point for a massless particle is

\begin{equation}
dt=\frac{dr}{\sqrt{\frac 43\Lambda M^2}}  \label{30}
\end{equation}
Since $\Lambda $ is of the order $10^{-56}cm^{-2}$, then it takes less than $%
\tau \equiv 10^5(\frac M{km})^{-1}$ years for a photon to escape from
horizon region of black hole i.e. $\rho \approx 2M$ to outside of few $M.$

Since the metric is nonsingular for $r\neq 0$ and there is no horizon, the
high potential regions are accessible. This huge amount of potential energy
can be converted to thermal or other physical forms of energy and providing
us with a new supermachinary for high energy astrophysics. This can make
drastic change in our understanding of the AGN phenomenon.

We should mention that the presented non-static metric and the
Schwarzschild-deSitter metric are transformable to each other by the
following coordinate transformation

\[
\rho =R(t)\;r\;\;\;\;\;\;\;\;\;\;\;\;\;\;\;\;\;\;\;\;\; 
\]

\begin{equation}
T=t+\frac{\dot{R}}R\int^\rho \frac{A(\tilde{\rho})\tilde{\rho}d\tilde{\rho}}{%
1-\frac{2M}{\tilde{\rho}}-\frac \Lambda 3\tilde{\rho}^2}  \label{31}
\end{equation}
where A is given by (26). For the special case of $M=0$, (31) reduces to

\begin{equation}
T=t-\frac 12\ln (1-\frac \Lambda 3\rho ^2)  \label{32}
\end{equation}
As it is evident from (32) this transformation does not preserve the
symmetries of space and this is the reason that why in
Schwarzschild-deSitter metric the spherical symmetry remains in the limit$%
M\rightarrow 0.$

In a recent work I have shown that there exist general spherically symmetric
solutions of Einstein vacuum field equations without $\Lambda $ [12]. These
solutions are free of intrinsic singularity and are analytic at $r=0$. We
anticipate that the combination of these two works will provide us the
ultimate nonsingular metrics.


\begin{thebibliography}{99}
\bibitem{1}  Leibundgut, B.,et al. $astro-ph/9812042$.

\bibitem{2}  Perlmutter, S. ,et al. $Nature$,391,51(1998).

\bibitem{3}  Schmidt, B.P. ,et al. $APJ$,507,46(1998).

\bibitem{4}  Granavich, P.M. ,et al. $APJ$,493,L53(1998).

\bibitem{5}  Riess, A.C. ,et al. $AJ$,116,1009(1998).

\bibitem{6}  Weinberg, S. $Rev.Mod.Phys.$ 61,1(1989).

\bibitem{7}  Ng, Y.J. $Int.J.Mod.Phys$. D1,145(1992).

\bibitem{8}  Razmi, H. $quant-ph/9811071$.

\bibitem{9}  Razmi, H.,Abbassi, Amir H. $quant-ph/9901022$.

\bibitem{10}  Rindler, W. $EssentialRelativity$ ,Springer-Verlag,p184,\
(1977).

\bibitem{11}  Abbassi, Amir H. ,Khosravi, Sh. $gr-qc/9812092$.

\bibitem{12}  Abbassi, Amir H. $gr-qc/9812081$.
\end{thebibliography}
\end{document}